\documentclass[preprint, showpacs,preprintnumbers,amsmath,amssymb,nofootinbib]{revtex4}
\usepackage{mathrsfs}
\usepackage{amsmath}
\usepackage{amsfonts}
\usepackage{latexsym}
\usepackage{amsfonts}
\usepackage{graphicx}
\usepackage{epsf}
\usepackage{dcolumn}
\usepackage{bm}

\newcommand{\bea}[1]{\begin{eqnarray}\label{#1}}
\newcommand{\eea}{\end{eqnarray}}

\def\gsim{ \lower .75ex \hbox{$\sim$} \llap{\raise .27ex \hbox{$>$}} }
\def\lsim{ \lower .75ex \hbox{$\sim$} \llap{\raise .27ex \hbox{$<$}} }

\begin{document}
 \title{The  dynamical behavior of $f(T)$ theory}

\author{Puxun Wu $^{1,2,3}$
and  Hongwei Yu  $^{2,1,3}$ \footnote{Corresponding author: hwyu@hunnu.edu.cn}
}
\address
{$^1$ Center of Nonlinear Science and Department of Physics, Ningbo
University,  Ningbo, Zhejiang, 315211 China\\
$^2$
Department of Physics and Institute of  Physics, Hunan Normal
University, Changsha, Hunan 410081, China \\
$^3$ Key Laboratory of Low Dimensional Quantum Structures and Quantum
Control of Ministry of Education, Hunan Normal University, Changsha,
Hunan 410081, China
}

\begin{abstract}
Recently, a new model obtained from generalizing  teleparallel
gravity, named  $f(T)$ theory,  is proposed to explain the present
cosmic accelerating expansion with no need of  dark energy. In this
letter, we analyze the dynamical property of this theory. For a
concrete power law model, we obtain that the dynamical system has a
stable de Sitter phase along with an unstable radiation dominated
phase and an unstable matter dominated one. We show that the
Universe can evolve from a radiation dominated era to a matter
dominated one, and finally enter an exponential expansion phase.

\end{abstract}
\maketitle

\section{Introduction}\label{sec1}
The models obtained from  modifying  general relativity theory are
viable candidates  to explain the current cosmic acceleration, which
was first discovered from the supernova observations
~\cite{Riess1998} and  then further confirmed by many other
cosmological tests, including the cosmic microwave background
radiation~\cite{Spergel2003} and the large scale
structure~\cite{Tegmark2004}, and so on. One such modification to
general relativity is the $f(R)$ theory~(see \cite{Felice2010} for
recent review), where the Ricci scalar  $R$ in the Einstein-Hilbert
action is generalized to  an arbitrary function $f$ of $R$.

Recently,  a new modified gravity to account for the cosmic
accelerating expansion, named $f(T)$ theory, is proposed by
extending the action of teleparallel gravity~\cite{Einstein1928,
Einstein1930} in analogy to the $f(R)$ theory, where $T$ is the
torsion scalar. The teleparallel theory of gravity is built on
teleparallel geometry, which was first introduced by Einstein to
unify gravity and electromagnetism~\cite{Einstein1928} and then was
revived as a geometrical alternative to the Riemannian geometry of
general relativity~\cite{Moller1961}. In teleparallel geometry,  the
Weitzenb\"{o}ck connection rather than the Levi-Civita connection is
used. As a result, the spacetime has only torsion and thus is
curvature-free.

It has been demonstrated that the $f(T)$ theory can  not only
explain the present cosmic acceleration with no need of dark
energy~\cite{Bengochea2009}, but also provide an alternative to
 inflation without an inflaton~\cite{Ferraro2007, Ferraro2008}.
 It therefore has attracted some attention recently. In this regard,
 Linder~\cite{Linder2010} proposed two new $f(T)$ models to explain
the present cosmic accelerating expansion and found that the $f(T)$
theory can unify a number of interesting extensions of gravity
beyond general relativity.  We performed  a statefinder diagnostic
to  these two models  and also placed observational constraints  on
them from the latest data~\cite{Wu2010}. More recently, a
reconstruction of $f(T)$ theory from the background expansion
history and  the $f(T)$ theory driven by scalar fields have been
studied~\cite{Myrzakulov2010, Yerzhanov2010}. In this letter, we plan
to  analyze the dynamical property of
$f(T)$ theory.

\section{The $f(T)$ theory}
In teleparallel gravity,  the dynamical object is the vierbein
$e^\mu_i$, which has the property,
\begin{eqnarray}
e^\mu_i e_\mu^j=\delta^j_i, \quad e^\mu_i e_\nu^i=\delta^\mu_\nu,
\end{eqnarray}
where   $e^i_\mu$ is the inverse matrix of vierbein,   $i$ is an
index running over $0,1,2,3$ for the tangent space of the manifold,
and $\mu$, also running over   $0,1,2,3$, is the coordinate index on
the manifold.  This vierbein relates with the metric through
\begin{eqnarray} g_{\mu\nu}=\eta_{ij}e^i_\mu e^j_\nu\;,
\end{eqnarray}
where $\eta_{ij}=diag(-1,1,1,1)$.

As mentioned in the previous section,  teleparallel gravity uses the
curvatureless Weitzenb\"{o}ck connection, which is defined as
\begin{eqnarray}  {\hat{\Gamma}}^\lambda_{\mu\nu}=e^\lambda_i\partial_\nu e^i_\mu=- e^i_\mu \partial_\nu e^\lambda_i\;.
\end{eqnarray}
From this Weitzenb\"{o}ck connection, one can introduce a non-null
torsion tensor $T^\sigma_{\;\;\mu\nu}$,
\begin{eqnarray}
T^\sigma_{\;\;\mu\nu}={\hat{\Gamma}}^\lambda_{\nu\mu}-{\hat{\Gamma}}^\lambda_{\mu\nu} \;.
\end{eqnarray}
The torsion scalar $T$ in the action of teleparallel gravity is then
given by
\begin{eqnarray}\label{ST}
T\equiv S^{\;\;\mu\nu}_\sigma T^\sigma_{\;\;\mu\nu}\;,
 \end{eqnarray}
where \begin{eqnarray}
S^{\;\;\mu\nu}_\sigma\equiv\frac{1}{2}(K^{\mu\nu}_{\;\;\;\;\sigma}+\delta^\mu_\sigma T^{\alpha \nu}_{\;\;\;\;\alpha}-\delta^\nu_\sigma T^{\alpha \mu}_{\;\;\;\;\alpha})\;,
\end{eqnarray}
 and $K^{\mu\nu}_{\;\;\;\;\sigma}$ is the contorsion tensor,
\begin{eqnarray}
 K^{\mu\nu}_{\;\;\;\;\sigma}=-\frac{1}{2}(T^{\mu\nu}_{\;\;\;\;\sigma}-T^{\nu\mu}_{\;\;\;\;\sigma}-T_{\sigma}^{\;\;\mu\nu}).
 \end{eqnarray}

For a flat homogeneous and isotropic Friedmann-Robertson-Walker
universe described by the metric $g_{\mu\nu}=diag(-1, a^2(t),
a^2(t), a^2(t))$ where $a$ is the scale factor, one has, from
Eq.~(\ref{ST}),
\begin{eqnarray}
T=-6 H^2\;,
\end{eqnarray}
with $H=\dot{a}a^{-1}$ being the Hubble parameter.

The action of $f(T)$ theory is obtained by replacing  $T$ in the
action of teleparallel gravity by  $T+f(T)$. Varying this action
with respect to the vierbein, we obtain the field equation of $f(T)$
gravity, which leads to
 the following modified
Friedmann equation
\begin{eqnarray}\label{MF}
H^2=\frac{8\pi G}{3}\rho-\frac{f}{6}-2H^2f_{T}\;,
\end{eqnarray}
\begin{eqnarray} \label{MF2}
(H^2)'=\frac{16\pi GP+6H^2+f+12H^2f_{T}}{24H^2f_{T T}-2-2f_{T}}\;,
\end{eqnarray}
where a prime denotes a derivative with respect to $\ln a$,
subscript $T$, a  derivative with respect to $T$, $\rho$ is the
energy density  and $P$ is the pressure. Here we assume that the
energy components in the Universe are matter and radiation, thus
\begin{eqnarray}
\rho=\rho_m+\rho_r, \quad P=\frac{1}{3}\rho_r\;,
\end{eqnarray}
where $\rho_m$ and $\rho_r$ represent the energy densities of matter
and radiation, respectively.

From Eqs.~(\ref{MF}, \ref{MF2}), we can define an effective dark
energy, whose  energy density and the equation of state can be
expressed, respectively, as,
\begin{eqnarray}
\rho_{eff}=\frac{1}{16\pi G}(-f + 2Tf_{T})
\end{eqnarray}
\begin{eqnarray}
w_{eff}=-\frac{f/T-f_{T}+2Tf_{TT}+\frac{1}{3} \frac{8\pi G \rho_r}{ 3 H^2}(f_{T}+2T f_{T T})}{(1+f_{T}+2Tf_{TT})(f/T-2f_{T})}\;.
\end{eqnarray}

\section{dynamical analysis}

In order to analyze the dynamics of a general $f(T)$ model, we
rewrite the equations of motion as a dynamical system with the
following dimensionless variables:
\begin{eqnarray}
x\equiv-\frac{f}{6H^2},\quad y\equiv \frac{T f_{T}}{3H^2},\quad z\equiv \Omega_r\equiv\frac{8\pi G \rho_r}{3H^2}\;,
\end{eqnarray}
where $\Omega_r$ is the dimensionless energy density parameter of
radiation. Using Eqs.~(\ref{MF}, \ref{MF2}) and the energy
conservation equations of matter and radiation, one can obtain
\begin{eqnarray}\label{x}
x'=-(2x+y)\frac{z+3-3x-3y}{2my-2+y}\;,
\end{eqnarray}
\begin{eqnarray}\label{y}
y'=2my\frac{z+3-3x-3y}{2my-2+y}\;,
\end{eqnarray}
\begin{eqnarray}\label{z}
z'=-4z-2z\frac{z+3-3x-3y}{2my-2+y}\;,
\end{eqnarray}
where a prime denotes a derivative with respect to $\ln a$ and
\begin{eqnarray}\label{dm}
m\equiv \frac{T f_{T T}}{f_{T}}\;.
\end{eqnarray}
Defining $r\equiv -2\frac{T f_{T}}{f}=\frac{y}{x}$, one can express
$T$ as a function of $y/x$ (or $r$). And then $m$ can be expressed
in terms of $y/x$. For example, the model, $f(T)= \alpha
[(-T)^p-\beta]^q$, yields $m(r)=(1-q)r/2q+p-1$.  Thus, for a given
form of $f(T)$, the dynamical system given in Eqs.~(\ref{x},
\ref{y}, \ref{z}) becomes autonomous.

Using $x$, $y$ and $z$, Eq.~(\ref{MF}) and $w_{eff}$ can be rewritten as
 \begin{eqnarray}\label{om}
 \Omega_m\equiv \frac{8\pi G \rho_m}{3H^2}=1-x-y-z\;,
 \end{eqnarray}
\begin{eqnarray}
w_{eff}=-\frac{x+y/2-my}{(1-y/2-my)(x+y)}\;,
\end{eqnarray}
where $\Omega_m$ is the dimensionless density parameter of matter.

In order to analyze the dynamical properties of system given in
Eqs.~(\ref{x}, \ref{y}, \ref{z}), we should firstly solve these
equations with $x'=0$, $y'=0$ and $z'=0$. Here, besides two isolated
critical points (denoted as Point A and Point B), we also get a
continuous line of critical points, which is called as Line C:
\begin{eqnarray}
Point\; A: \quad
   x_c=0,\; y_c=0, \; z_c=1\;,
\end{eqnarray}
\begin{eqnarray}
Point\; B: \quad
   x_c=0,\; y_c=0, \; z_c=0\;,
\end{eqnarray}
\begin{eqnarray}
Line\; C: \quad
   x_c=1-y_c, \; z_c=0\;.
\end{eqnarray}
One can see that Line C is a straight line in the phase space.

$\bullet$ Point A: radiation dominated point

At this critical point, we have
\begin{eqnarray}
\Omega_{r}=1\;,
\end{eqnarray}
which corresponds to a radiation dominated phase. Now we examine the
stability of this point, which is determined by the eigenvalues of
the linearized system. After some calculations, we find the
eigenvalues at Point A,
\begin{eqnarray}
1,\quad 2 (1 - m \pm \sqrt{1 + 2 m + m^2 - 2 m'})\;,
\end{eqnarray}
where $m'=dm/dr$.  This critical point is unstable  because there is
a positive eigenvalue.

$\bullet$ Point B: matter dominated point

Using Eq.~(\ref{om}), one has
\begin{eqnarray}\Omega_{m}=1\;,
\end{eqnarray}
at this critical point. Thus, it represents  a matted dominated era.
Through the same calculation as that in Point A, we obtain that the
eigenvalues of the linearized system at this point are,
\begin{eqnarray}
-1,\quad 2 - 2m \pm 2\sqrt{1 + 2 m + m^2 - 2 m'}\;.
\end{eqnarray}
For a successful cosmological scenario, this Point must be unstable
so that the Universes can exit from the the matter dominated era.
This means that the real part of one of the eigenvalues in the above
expression must be positive. This may be possible for  a given
$f(T)$ model as long as the model parameters satisfy certain
conditions.

$\bullet$ Line C: effective dark energy  dominated era

At this critical line,  we have $\Omega_{m}=0$,  $\Omega_{r}=0$ and
\begin{eqnarray}
 w_{eff}=-1\;.
\end{eqnarray}
Substituting Eq.~(\ref{MF}) into (\ref{MF2}) and considering $\rho=0$ and $p=0$, one can find easily that
\begin{eqnarray}
(H^2)'=0\;.
\end{eqnarray}
Thus, Line C corresponds to a de Sitter phase, if $H\neq 0$. The
eigenvalues of the linearized system at this critical line are,
\begin{eqnarray}
-4,\quad 0,\quad -3\;.
\end{eqnarray}
It is easy to see that Line C is always stable. Thus, for a given
$f(T)$ model, the Universe finally enters a de Sitter phase.

Now we consider a concrete   power law model given in~\cite{Bengochea2009, Linder2010}:
\begin{eqnarray}\label{PL} f(T)=\alpha (-T)^n\;,\end{eqnarray} where $\alpha$ and $n$ are two model parameters.
In Refs.~\cite{Linder2010, Wu2010}, it has been pointed out that
this model has the same background evolution equation as some
phenomenological models~\cite{Dvali2003, Chung2000} and  it reduces
to the $\Lambda$CDM model when $n=0$,   and to the DGP
model~\cite{Dvali2000} when $n=1/2$. When $n=1$, the Friedmann
equation (Eq.~(\ref{MF})) can be rewritten as $H^2=\frac{8\pi
G}{3(1-\alpha)}\rho$, which is the same as that of a standard cold
dark matter (SCDM) model if we rescale the Newton's constant as
$G\rightarrow G/(1-\alpha)$. Thus, next,  we will focus our
attention on the case of $n\neq 1$. Let us note that,  in order to
be consistent with the present observational results, it is required
that $|n|\ll 1$~\cite{Bengochea2009, Linder2010, Wu2010}.

\begin{figure}[htbp]
 \centering
\includegraphics[width=6cm]{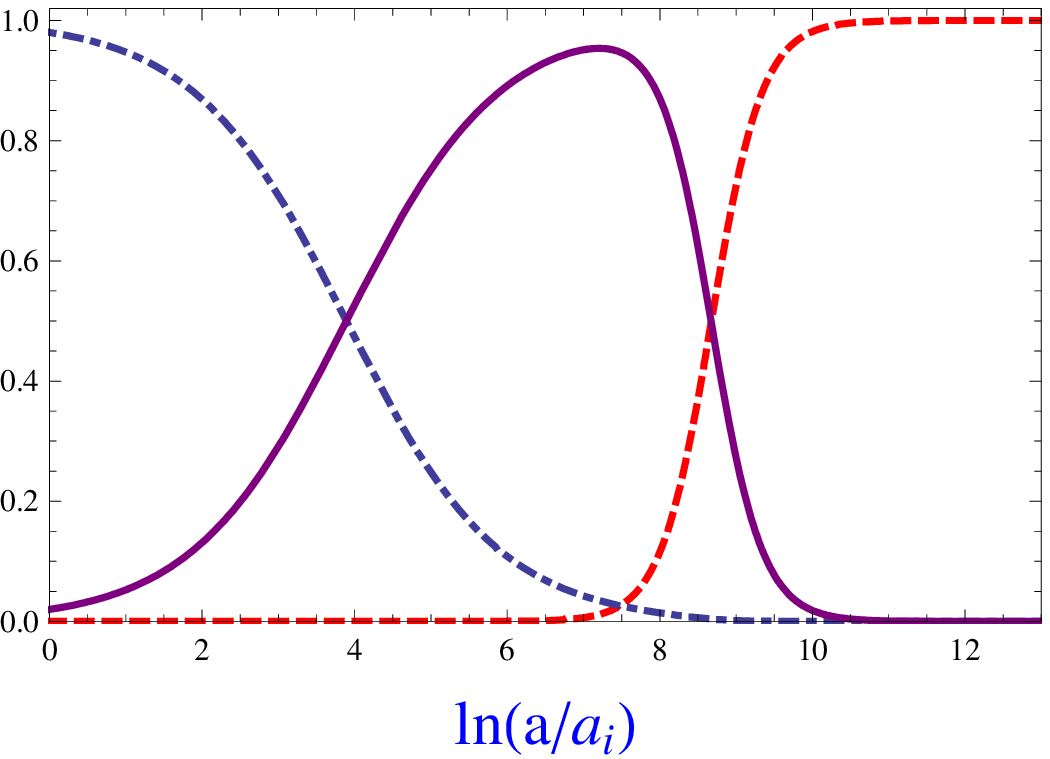}\quad\includegraphics[width=6cm]{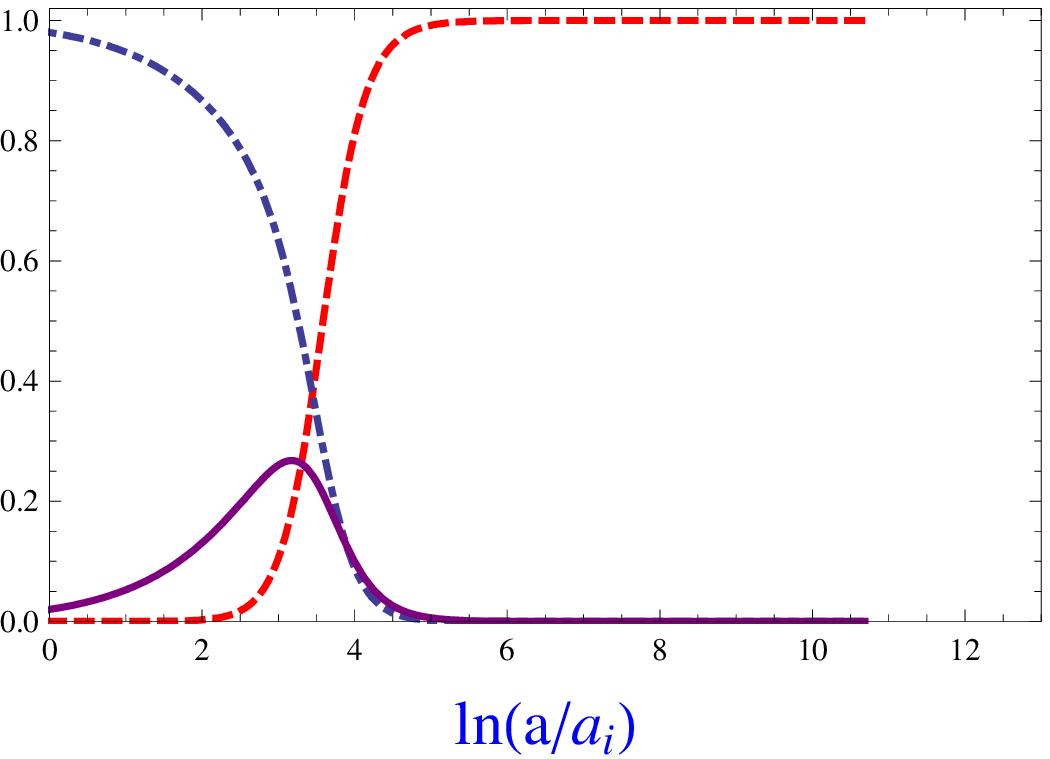}
\caption{\label{Fig1}  The cosmic evolution for the case of power
law model with $n=0.2$.  $a_i$ is the initial value of the scalar
factor.  The dot-dashed, solid, and dashed lines correspond to the
evolutionary of curves of $\Omega_{r}$, $\Omega_m$ and the
dimensionless density parameter of the effective dark energy,
respectively. In left panel the initial conditions are set as
$x_i=y_i=10^{-13}$ and $z_i=0.98$, while in right panel they are
$x_i=y_i=10^{-5}$ and $z_i=0.98$.}
 \end{figure}

Substituting Eq.~(\ref{PL})  into Eq.~(\ref{MF}), one can show that,
when   $n\neq 1$, the case   $\rho=0$ (Line C) gives that the Hubble
parameter is a non-zero constant, which corresponds to a de Sitter
phase.  Using Eq.~(\ref{dm}), we have $m=-1+n$ and $m'=0$. The
eigenvalues at critical point B becomes
\begin{eqnarray} -1,\quad 2, \quad -2n,\end{eqnarray} which means
that critical point B is always unstable and its stability is
independent of the value of model parameter $n$.  Therefore, for a
power law model,  we find that, if $n\neq 1$, the Universe is able
to evolve from a radiation dominated era to a matter dominated one,
and finally enter an exponential expansion phase.  In
Fig.~(\ref{Fig1}), we show the cosmic evolution with different
initial conditions. It is easy to see that $x_i$ and $y_i$ ($f/T$
and $-2f_T$ at $a_i$, where $a_i$ is the initial value of scale
factor) must be very very small to have a long enough period of
radiation domination give the correct primordial nucleosynthesis and
radiation-matter equality, and to ensure the appearance of  a matter
dominated phase, otherwise the Universe has unusual early behavior
and evolves directly from radiation dominated phase to a de-Sitter
one. Thus, for the power law $f(T)$ model,  the conditions for the
Universe to evolve to a cosmic accelerating expansion and have usual
early behavior are $x_i\ll 1$, $y_i \ll 1$ and $n\neq 1$. Note,
however, that in order to satisfy the current observations
constraints, $|n|\ll1$ is still required~\cite{Bengochea2009,
Linder2010, Wu2010}.

\section{Conclusion}
The $f(T)$ theory, obtained from generalizing  teleparallel gravity,
is a new modified gravity capable of accounting for the present
cosmic accelerating expansion  with no need of dark energy. In this
Letter,  we analyze the dynamical behavior of the $f(T)$ theory by
assuming the existence of matter and radiation in our Universe. Two
critical points (Point A and Point B), corresponding to a matter
dominated phase and a radiation dominated one, respectively, and a
critical line (Line C), corresponding to an effective dark energy
dominated era, are found. We find that both Point A and Point B are
unstable while Line C is always stable. Thus, the Universe can
finally enter a de Sitter expansion phase,  if $H$ is nonzero at the
critical Line C. For a power law model, the case $n\neq 1$ is
considered since $n=1$ corresponds to  a SCDM model if we rescale
the Newton's constant as $G\rightarrow G/(1-\alpha)$.  The results
show that, if $n\neq 1$, the final state of our Universe in the
$f(T)$ theory is an exponential expansion since $H$ is a nonzero
constant at Line C.  In addition, we find that, to obtain the usual
early universe behavior, it is required that $x_i$ and $y_i$ ($f/T$
and $-2f_T$ at $a_i$) should be very very small. Thus, for the power
law model,  the conditions to have a successful cosmological
scenario are that $x_i\ll 1$, $y_i \ll 1$ and $n\neq 1$. But,
according to the results obtained in Refs.~\cite{Bengochea2009,
Linder2010, Wu2010}, $|n|\ll1$ is required  in order to be
consistent with current observations.

\begin{acknowledgments}

This work was supported in part by the National Natural Science
Foundation of China under Grants Nos. 10775050, 10705055 and
10935013, the SRFDP under Grant No. 20070542002,  the FANEDD under
Grant No. 200922, the National Basic Research Program of China under
Grant No. 2010CB832803, the NCET under Grant No. 09-0144, the PCSIRT under Grant No. IRT0964, the Programme for the Key Discipline in
Hunan Province, and K.C. Wong Magna Fund in Ningbo University.

\end{acknowledgments}

\end{document}